\documentclass[%
 reprint,
nofootinbib,
 aps,
prd,
]{revtex4-2}

\usepackage{graphicx}
\usepackage{dcolumn}
\usepackage{bm}
\usepackage{booktabs}
\usepackage{mathtools}
\usepackage{orcidlink}
\usepackage{hyperref}

\begin{document}

\title{Compact Dark Matter with Globular Cluster Pulsars}

\author{Soobeom Chung\,\orcidlink{0009-0006-7731-5395}}
\email{soobeomchung@ufl.edu}
\affiliation{Institute for Fundamental Theory, Physics Department, University of Florida, Gainesville, FL 32611, USA}

\author{Jeff A. Dror\,\orcidlink{0000-0003-0110-6184}}
\email{jeffdror@ufl.edu}
\affiliation{Institute for Fundamental Theory, Physics Department, University of Florida, Gainesville, FL 32611, USA}

\date{\today}
\begin{abstract}
Pulsars are exceptionally stable clocks, and a compact dark-matter object passing near the line of sight to a pulsar delays its pulses through the Shapiro effect. The number of detectable transits is set by the integral over the dark-matter density along the line of sight connecting the Earth and the pulsar. Established pulsars in pulsar timing arrays are typically within one kiloparsec of the Earth and hence sit in a dark-matter density comparable to our local value. In this work, we explore globular clusters in the inner Galaxy as targets to search for compact dark matter. These high-pulsar-density environments can be cheaply observed with a single telescope beam and enjoy a large dark-matter crossing rate. Additionally, since the recently detected nanohertz gravitational-wave background is nearly common to the cluster's pulsars, a multi-pulsar analysis can remove much of its contribution. Ranking known clusters, we find that Terzan 5, with 49 pulsars, offers the best reach. A forty-year campaign of weekly four-hour visits at the sensitivity envisioned for the second phase of the Square Kilometre Array (SKA) would probe compact objects making up $6\%$ of the dark matter at $0.02\,M_\odot$. Finally, we estimate the foreground from stars and stellar remnants in Terzan 5 and along the sightline. It is subdominant near the most sensitive low-mass part of the SKA benchmark, but becomes important over parts of the stellar mass range.
\end{abstract}
\maketitle
\section{Introduction}
\label{sec:intro}
The nature of dark matter remains one of the most important open questions in physics. If some of it is clumped into compact objects, such as primordial black holes (PBHs)~\cite{Carr:2016drx,Carr:2020gox,Carr:2020xqk, Green:2020jor,Escriva:2022duf} or dense subhalos, they can be detected directly through Galactic gravitational probes. Millisecond pulsars (MSPs) are among the most stable clocks in nature, so small shifts in the arrival times of their pulses reveal the gravitational potential between the pulsar and the Earth. In particular, a compact object passing near the line of sight delays the pulses as they cross its gravitational potential, producing a Shapiro delay. Searches for dark matter with pulsar timing were proposed in Refs.~\cite{Seto:2007kj,Siegel:2007fz, Baghram:2011is, Kashiyama:2012qz, Khmelnitsky:2013lxt, Schutz:2016khr, Kim:2023kyy}, and the signals of PBHs and extended subhalos were studied in detail in Refs.~\cite{Dror:2019twh, Ramani:2020hdo, Lee:2021zqw}. These works showed that pulsar timing complements gravitational lensing, remaining sensitive even to objects too diffuse to lens light appreciably.

The Shapiro delay accumulates along the entire photon path, so a compact object anywhere between the Earth and the pulsar can produce a signal. The number of objects that pass close enough to the sightline during an observing campaign is set by the dark-matter column, the total dark-matter mass per unit area along the sightline:
\begin{equation}
\Sigma_{\rm DM} = \int_0^{D_{\rm P}} \rho_{\rm DM}(\ell)\, d\ell,
\label{eq:column-def}
\end{equation}
where $\rho_{\rm DM}(\ell)$ is the dark-matter density a distance $\ell$ from the Earth along the line of sight and $D_{\rm P}$ is the distance to the pulsar. For compact objects of mass $M$ making up a fraction $f$ of the dark matter, the number of objects per unit area projected onto the sky is $f\,\Sigma_{\rm DM}/M$. As a result, the reach on the dark-matter fraction scales inversely with the column, $f \propto 1/\Sigma_{\rm DM}$, making the column a key figure of merit for a Shapiro search. Previous studies, however, relied on the pulsars monitored by pulsar timing arrays (PTAs)~\cite{Dror:2019twh, Schutz:2016khr}, which are selected for their timing stability in the search for gravitational waves rather than for the dark matter along their sightlines. The column is largest toward the inner Galaxy, so pulsars there have far more dark matter along their sightlines.

In this work, we explore pulsars in globular clusters as targets for dark-matter detection. A single cluster can host dozens of MSPs within a few parsecs, so one telescope pointing times many sightlines through the same large column. Terzan 5 is the most extreme example, with 49 known pulsars, the most of any globular cluster~\cite{Padmanabh2024}. Timing of these pulsars over a 10-yr baseline has already been used to measure the mass distribution of the cluster~\cite{Prager2017}. However, the Terzan 5 pulsars are faint, leading to published timing residuals at the $10$--$100~\mu$s level~\cite{Padmanabh2024,Cadelano2018,Corcoran2026, Andersen2018, Ridolfi2021}, two to three orders of magnitude larger than those of the best PTA pulsars. Nevertheless, we find that timing these pulsars with the Square Kilometre Array would bring compact dark matter within reach.

A second obstacle is the nanohertz stochastic gravitational-wave background (GWB). PTAs have recently found evidence for a GWB with the Hellings--Downs angular correlations~\cite{HellingsDowns1983} expected of an isotropic background~\cite{NANOGrav2023,EPTA2023}. This acts as a background for a dark-matter search, suppressing the sensitivity of existing arrays~\cite{Cherukupalli:2026cda,Foster:2026kfg}. However, the pulsars of a globular cluster are separated by less than the wavelength of a nanohertz gravitational wave, so the background is nearly identical across the cluster. In contrast, a transiting object delays the pulses along a single sightline. An analysis exploiting this correlation can therefore remove the common background, recovering much of the sensitivity lost to the GWB.

Finally, the Shapiro delay depends only on the mass and trajectory of the transiting object, so an ordinary star produces the same signal as a dark object of the same mass. Stars and stellar remnants in Terzan 5 and in the disk and bulge of the Milky Way all cross the sightlines to the cluster. We quantify this foreground through the dark-matter fraction that would produce the same number of transits, which we refer to as the compact-baryon floor.

In Sec.~\ref{sec:signal}, we review the Shapiro signal of a transiting compact object and derive the dependence of the reach on the dark-matter column. In Sec.~\ref{sec:column}, we compute the column toward the globular clusters that host pulsars and rank them as targets. In Sec.~\ref{sec:baryon}, we construct the compact-baryon floor from the stars and stellar remnants of Terzan 5 and the Milky Way. In Sec.~\ref{sec:results}, we present the projected sensitivity.

\section{The Shapiro Signal and the Line-of-Sight Column}
\label{sec:signal}
A compact object passing near the line of sight to a pulsar delays the arrival times of its pulses. As the object moves, the delay changes with time, so the observed spin frequency $\nu$ of the pulsar shifts. For an object of mass $M$ moving with transverse speed $v$ relative to the line of sight at impact parameter $b$, and reaching closest approach at time $t_0$, the shift is~\cite{Dror:2019twh}
\begin{equation}
\frac{\delta\nu}{\nu} = \frac{4GM\,v^2(t-t_0)}{b^2 + v^2(t-t_0)^2}.
\label{eq:freq-shift}
\end{equation}
Except for compact objects very close to the Earth or pulsar, the shift depends only on the motion of the object transverse to the line of sight and not on its position along it. Integrating Eq.~\eqref{eq:freq-shift} over time gives the timing residual,
\begin{equation}
R(t) = -2GM \ln\!\big[\tau^2 + (t-t_0)^2\big],
\label{eq:shapiro-residual}
\end{equation}
up to an unobservable constant, where $\tau \equiv b/v$ is the duration of the transit.

Only part of this residual is observable. The arrival times of each pulsar are first fit with a timing model, and only the residuals of that fit are searched, so any part of a transit that the model can mimic is removed with it~\cite{vanHaasterenLevin2013}. The model predicts the arrival times from the astrometry of the pulsar and its spin. The astrometric parameters (position, proper motion, and parallax) produce delays that vary with the yearly motion of the Earth, which does not resemble a transit lasting months to years, so they do not absorb it.\footnote{For a binary pulsar the model also includes the orbital parameters, whose delays vary with the orbital period of hours to days and so do not absorb a transit either; we neglect this complication.} The spin parameters do. The pulsar phase is a polynomial in time, $\phi(t) = \phi_0 + \nu t + \frac{1}{2}\dot\nu t^2 + \dots$, and for a pulsar outside a cluster the fit typically stops at $\dot\nu$, which absorbs the intrinsic spin-down. For a pulsar in a globular cluster the fit must go further. The line-of-sight acceleration of the pulsar in the potential of the cluster, and that from nearby stars, changes over the campaign, which adds measurable second and third derivatives of the spin frequency, $\ddot\nu$ and $\dddot\nu$: of the 36 Terzan 5 pulsars timed in Ref.~\cite{Prager2017}, 34 have a measured second derivative and 13 a third or higher, and a third derivative is marginally detected in 47~Tuc as well~\cite{Freire:2017mgu}. The model therefore absorbs any component of the residual that resembles a polynomial of degree up to four in time.

A transit is detected through the part of $R$ orthogonal to the fitted polynomials, $R_\perp$. For white timing noise its matched-filter significance is
\begin{equation}
\begin{aligned}
{\rm SNR}^2 &= \sum_i \frac{R_\perp(t_i)^2}{t_{\rm RMS}^2} \\
&= C\,\frac{(GM)^2\,T}{t_{\rm RMS}^2\,\Delta t},
\end{aligned}
\label{eq:snr-c}
\end{equation}
where the sum runs over the observations, $t_{\rm RMS}$ is the root-mean-square timing residual of a single observation, $\Delta t$ is the time between observations, $T$ is the length of the campaign, and $C \equiv \langle R_\perp^2\rangle/(GM)^2$ is the mean square of the post-fit residual over the campaign.

The coefficient $C$ depends only on $t_0/T$ and $\tau/T$; App.~\ref{app:timing-fit} describes how we compute it. For a short transit at the middle of the campaign, $\tau \ll T$ and no timing-model fit, $C = 32$, which reproduces the significance of Ref.~\cite{Dror:2019twh},
\begin{equation}
{\rm SNR} \simeq GM \sqrt{\frac{32\, T}{t_{\rm RMS}^2 \Delta t}}.
\label{eq:snr-dyn}
\end{equation}
We require ${\rm SNR} > 4$ for a detection, leading to a no-fit detection threshold
\begin{equation}
M_{\rm th} = \frac{t_{\rm RMS}}{\sqrt{2}\, G}\sqrt{\frac{\Delta t}{T}}.
\label{eq:mth}
\end{equation}
It is convenient to measure the mass in units of $M_{\rm th}$: a general transit is detectable if
\begin{equation}
C(t_0,\tau) > 32\,\bigg(\frac{M_{\rm th}}{M}\bigg)^2 .
\label{eq:detect}
\end{equation}
Without the fit, every short transit of an object heavier than $M_{\rm th}$ is detectable; after the fit, the longer transits require heavier objects.

In the plane transverse to the line of sight, the position of an object at the start of the campaign is specified by its distance $b$ from the sightline perpendicular to its velocity and its distance $x$ from closest approach along its velocity. Both are fixed by times, $b = v\tau$ and $x = v t_0$, so the area element is $db\,dx = v^2\,d\tau\,dt_0$. The number of objects per unit area projected onto the sky is $f\,\Sigma_{\rm DM}/M$, so the expected number of detectable transits is
\begin{equation}
\langle N \rangle = \frac{f\,\Sigma_{\rm DM}}{M}\,A(M),
\label{eq:nexp}
\end{equation}
where the detectable area,
\begin{equation}
A(M) = 2v^2 \int_0^\infty d\tau \int_{-\infty}^{\infty} dt_0\;
\Theta\!\bigg[C(t_0,\tau) - 32\,\frac{M_{\rm th}^2}{M^2}\bigg],
\label{eq:area}
\end{equation}
counts every trajectory that satisfies Eq.~\eqref{eq:detect}, $\Theta$ is the Heaviside step function, and the factor of 2 counts both sides of the sightline. For a population with a distribution of speeds, $v^2$ is replaced by its mean square, $\langle v^2\rangle$. The simplified limit of Ref.~\cite{Dror:2019twh} assumed that, for $M > M_{\rm th}$, the argument of the Heaviside step function is positive for $\tau < t_0 < T - \tau$, leading to $A = (vT)^2/2$.

The probability of observing no transit is $e^{-\langle N \rangle}$, which falls to $10\%$ when $\langle N \rangle = \ln 10$. In the absence of a detection, the fraction is therefore excluded above
\begin{equation}
f = \frac{\ln 10\; M}{\Sigma_{\rm DM}\, A(M)}.
\label{eq:schematic-reach}
\end{equation}

For a cluster of $N_{\rm P}$ pulsars, the sightlines are farther apart than the distance $vT$ that an object travels during the campaign, except for objects passing within ${\cal O}(10^2~{\rm pc})$ of the Earth.\footnote{Closer to the Earth the sightlines converge and an object can cross several of them within a campaign; we neglect this contribution.} As such, the detectable areas simply add: $A(M)$ in Eq.~\eqref{eq:schematic-reach} is replaced by the sum of the single-sightline areas of Eq.~\eqref{eq:area}, each evaluated with its pulsar's own $M_{\rm th}$ and timing model.

As a benchmark, consider the simplified $A = (vT)^2/2$ limit. Combining Eq.~\eqref{eq:mth} with Eq.~\eqref{eq:schematic-reach} gives the reach at $M \simeq M_{\rm th}$,
\begin{equation}
f = \sqrt{2}\,\ln 10\,
\frac{t_{\rm RMS}\sqrt{\Delta t/T}}
{G\,(vT)^2\,N_{\rm P}\,\Sigma_{\rm DM}}.
\label{eq:best-reach-general}
\end{equation}
While this estimate is parametrically accurate, the fit moves the best reach to $M \simeq 10\,M_{\rm th}$ and weakens it by a factor of about 30, as follows from the detectable area shown in Fig.~\ref{fig:area}. Including this factor, for benchmark values relevant for Terzan 5, with $N_{\rm P} = 49$, a column $\Sigma_{\rm DM} \simeq 200\,M_\odot\,{\rm pc}^{-2}$, computed in Sec.~\ref{sec:column}, and $v = 250~{\rm km\,s^{-1}}$, we roughly get
\begin{equation}
f \simeq 0.4
\bigg(\frac{t_{\rm RMS}}{1\ \mu{\rm s}}\bigg)
\bigg(\frac{\Delta t}{1\ {\rm week}}\bigg)^{1/2}
\bigg(\frac{40\ {\rm yr}}{T}\bigg)^{5/2},
\label{eq:best-reach}
\end{equation}
at a mass $M \simeq 10\,M_{\rm th} \simeq 0.03\,M_\odot$ for the benchmark values.

\section{Inner-Galaxy Globular Clusters as Targets}
\label{sec:column}
In this section, we compute the line-of-sight column for different globular clusters in the Milky Way containing pulsars and rank them by the figure of merit $N_{\rm P}\Sigma_{\rm DM}$, where $N_{\rm P}$ is the number of pulsars in the cluster.

\subsection{The dark-matter column}
\label{sec:column-model}
Globular clusters are not expected to carry a significant amount of dark matter~\cite{Moore:1995pb,Conroy:2010bs}. As such, when evaluating the dark-matter column, we focus on the contribution from the Milky Way halo, which we model with a Navarro--Frenk--White (NFW) profile~\cite{Navarro1996},
\begin{equation}
\rho_{\rm MW}(r) = \frac{\rho_s}{(r/r_s)\,(1 + r/r_s)^2},
\label{eq:nfw}
\end{equation}
with a scale radius $r_s = 20$~kpc~\cite{McMillan2017}. The scale density $\rho_s$ is fixed by requiring the local dark-matter density to be $\rho_\odot = 0.01\,M_\odot\,{\rm pc}^{-3}$ at the galactocentric radius of the Sun, $R_0 = 8.2$~kpc~\cite{McMillan2017}. Along a sightline toward a source at heliocentric distance $d$ and angle $\psi$ from the Galactic Center, a point a distance $\ell$ from the Sun lies at galactocentric radius
\begin{equation}
r(\ell) = \sqrt{R_0^2 + \ell^2 - 2R_0 \ell \cos\psi},
\label{eq:rx}
\end{equation}
so the column is
\begin{equation}
\Sigma_{\rm DM}(d,\psi) = \int_0^d d\ell\, \rho_{\rm MW}\big(r(\ell)\big).
\end{equation}

It is illustrative to compare this column to that of a 1~kpc sightline averaged over all directions on the sky, $\Sigma_{1\,\rm kpc}$, typical of a nearby disk pulsar. Sightlines through the inner Galaxy can have $\Sigma_{\rm DM} \gg \Sigma_{1\,\rm kpc}$; Terzan 5, for example, reaches $18\,\Sigma_{1\,\rm kpc}$, as listed in Table~\ref{tab:targets}.

\begin{table*}[tbp]
\begin{minipage}[t]{0.48\textwidth}
\vspace{0pt}
\begin{ruledtabular}
\begin{tabular}{lcccc}
Cluster & $N_{\rm P}^{\rm timed}/N_{\rm P}$ & $d$ [kpc] & $\Sigma_{\rm DM}/\Sigma_{1\,\rm kpc}$ & $N_{\rm P}\Sigma_{\rm DM}/\Sigma_{1\,\rm kpc}$ \\
\hline
Terzan 5 & 16/49 & 6.6 & 18 & 890 \\ \hline
NGC 6441 & 4/11 & 12.7 & 61 & 670 \\ \hline
NGC 6517 & 14/21 & 9.2 & 25 & 530 \\ \hline
NGC 6624 & 4/14 & 8.0 & 31 & 430 \\ \hline
NGC 6388 & 0/11 & 11.2 & 38 & 410 \\ \hline
NGC 6440 & 8/11 & 8.3 & 33 & 360 \\ \hline
M70 & 0/6 & 9.4 & 35 & 210 \\ \hline
M80 & 1/7 & 10.3 & 29 & 200 \\ \hline
M62 & 9/10 & 6.4 & 16 & 160 \\ \hline
NGC 6522 & 2/6 & 7.3 & 25 & 150 \\ \hline
M28 & 4/14 & 5.4 & 11 & 150 \\ \hline
$\omega$ Cen & 8/19 & 5.4 & 7 & 130 \\ \hline
Terzan 1 & 7/8 & 5.7 & 12 & 99 \\
\end{tabular}
\end{ruledtabular}
\end{minipage}%
\begin{minipage}[t]{0.52\textwidth}
\vspace{0pt}

\vspace*{4.56pt}\includegraphics{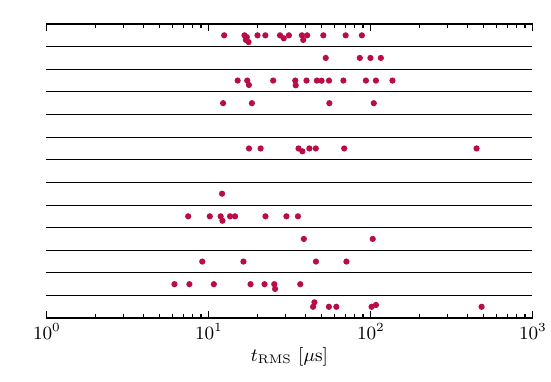}
\end{minipage}
\caption{{\bf Left:} Globular clusters hosting a pulsar array ($N_{\rm P} \geq 5$) in the inner Galaxy and, for comparison, $\omega$~Cen, discussed in Sec.~\ref{sec:ranking}: number of pulsars with a published timing residual, $N_{\rm P}^{\rm timed}$, out of the number known, $N_{\rm P}$~\cite{FreireGCpsr}, heliocentric distance~\cite{Baumgardt:2021gcd} (updated online catalog), line-of-sight column gain $\Sigma_{\rm DM}/\Sigma_{1\,\rm kpc}$ (NFW Milky Way halo), and the reach figure of merit $N_{\rm P}\Sigma_{\rm DM}$. Rows are ordered by $N_{\rm P}\Sigma_{\rm DM}$, since the single-event Shapiro reach scales as $f \propto 1/(N_{\rm P}\Sigma_{\rm DM})$: Terzan 5's large array places it first despite a lower column than most of the other globular clusters listed. {\bf Right:} timing residual of each pulsar with a published timing solution, on the row of its cluster~\cite{Freire2008,Vleeschower2022,Yin2024,Lynch2012,Perera2017,Ridolfi2021,Das2025,Zhang2020,Abbate2023,Douglas2022,Corcoran2026,Verbiest2009,Vleeschower2024,Andersen2018,ColomBernadich2026,Singleton2024,Cadelano2018,Padmanabh2024,Rosenthal:2024zzg}.}
\label{tab:targets}
\end{table*}

\subsection{Ranking the targets}
\label{sec:ranking}

Table~\ref{tab:targets} ({\bf Left}) ranks the inner-Galaxy globular clusters that currently host a pulsar array by the reach figure of merit $N_{\rm P}\Sigma_{\rm DM}$. Because $f \propto 1/(N_{\rm P}\Sigma_{\rm DM})$, the large pulsar population of Terzan 5 outweighs its lower column and gives the best reach. Twelve inner-Galaxy clusters presently host five or more known pulsars~\cite{FreireGCpsr}; the other inner-Galaxy clusters carry comparable columns but host too few known pulsars to form an array.

Sightlines with comparable or larger columns exist outside globular clusters, but none yet offers a timeable array. The sightline toward the Galactic Center carries the largest column, but it threads a strongly scattering region of the Galaxy~\cite{Macquart2010}, where only a few heavily scattered pulsars are known~\cite{Spitler2014,Dexter_2017}. The Sagittarius dwarf galaxy ($\Sigma_{\rm DM}/\Sigma_{1\,\rm kpc} \sim 50$) hosts the massive cluster M54, in which no pulsars have yet been timed~\cite{BiggsLyne1996,Evans2023Sgr}. The Magellanic Clouds ($\Sigma_{\rm DM}/\Sigma_{1\,\rm kpc} \sim 14$--$25$)~\cite{Prayag2024LMC,Carli2024SMCI,Carli2024SMCII, Prayag2026LMC} and M31 ($\Sigma_{\rm DM}/\Sigma_{1\,\rm kpc} \sim 26$)~\cite{vanLeeuwen2020M31, RubioHerrera2013M31} add large columns through their own halos, but no MSP has been timed in any of them: at 50--800~kpc the flux-limited timing precision degrades as $t_{\rm RMS} \propto d^2$, and with it the threshold mass $M_{\rm th} \propto t_{\rm RMS}$. The Magellanic Clouds are the most promising of these, because their high Galactic latitude avoids the inner-Galaxy scattering wall; should the Square Kilometre Array time even a modest array of bright Magellanic MSPs, the Large Magellanic Cloud would be comparable to the inner-Galaxy clusters. High columns are thus not unique to globular clusters. What sets clusters apart is the many co-located pulsars behind a single column, whose common noise can be removed. The entire cluster also lies in one beam, so a dedicated campaign is relatively inexpensive: all 49 pulsars of Terzan 5 are timed simultaneously in every visit.

$\omega$~Cen is the most massive globular cluster of the Milky Way, with $3.6\times 10^6\,M_\odot$~\cite{Baumgardt:2018pyl}, and is widely regarded as the stripped nucleus of an accreted dwarf galaxy~\cite{Bekki:2003qw}. It hosts 19 pulsars~\cite{Chen_2023,ColomBernadich2026}. Its sightline lies $15^\circ$ above the Galactic plane and misses the inner Galaxy, so the Milky Way column is only $7\,\Sigma_{1\,\rm kpc}$ and it ranks near the bottom of Table~\ref{tab:targets}. Nevertheless, measurements of stellar kinematics prefer a non-luminous component in addition to the stars, with a mass of up to $10^6\,M_\odot$ inside the half-light radius~\cite{Evans:2021bsh}, suggesting that $\omega$~Cen could have its own dark-matter profile.\footnote{This property has motivated searches and forecasts for radio signals of dark-matter annihilation and axion conversion in the cluster~\cite{Kar:2020coz, Wang:2021hfb, Wang_2024}.}

That said, compact dark objects bound to the cluster share the bulk motion of the pulsars, so their relative transverse speed is set by the internal velocity dispersion of the cluster and is of order $35~{\rm km\,s^{-1}}$~\cite{Baumgardt:2018pyl} rather than $250~{\rm km\,s^{-1}}$. Since the number of transits scales as the square of the relative speed, this significantly diminishes the detection prospects of any bound component. Estimating the column of such a bound component in $\omega$~Cen, we do not find the prospects for detection of compact dark matter promising unless one makes overly optimistic assumptions about the dark-matter profile, so we do not consider it further.

\subsection{Timing precision}
\label{sec:radiometer}
The pulsars along high-column inner-Galaxy sightlines are distant and therefore faint: the published Terzan 5 residuals span $\sim 10$--$100\ \mu$s~\cite{Padmanabh2024,Cadelano2018,Corcoran2026}, two to three orders of magnitude above the $\sim 50$~ns of the best pulsars outside clusters (see Table~\ref{tab:targets} ({\bf Right}) for the residuals of the most promising cluster targets). To project future capabilities, we first estimate how white noise, the leading noise source today, will fall with the sensitivity of the telescope. We then turn to the noise that our forecasts leave out: red noise intrinsic to the pulsars, chromatic noise from the interstellar medium, and the timing structure specific to a dense cluster, which the timing-model fit must absorb.

For a pulsar limited by radiometer noise, the residual of a visit is the effective pulse width divided by the signal-to-noise ratio of the folded profile~\cite{Cordes:2010fh}. This gives
\begin{equation}
t_{\rm RMS} \simeq \frac{T_{\rm sys}}{A_{\rm eff}}\,\frac{\sqrt{2}\,W^{3/2}}{S\,P^{1/2}\,(B\, t_{\rm obs})^{1/2}},
\label{eq:radiometer}
\end{equation}
where $A_{\rm eff}/T_{\rm sys}$ is the sensitivity of the telescope (effective area over system temperature), $S$ the flux density received from the pulsar (the power per unit collecting area and per unit frequency at the Earth, averaged over the pulse period and measured in Jy $= 10^{-26}\,{\rm W\,m^{-2}\,Hz^{-1}}$), $P$ its period, $W \ll P$ the duration of its pulse, $B$ the bandwidth, and $t_{\rm obs}$ the duration of the visit, with both polarizations summed.\footnote{Fitting the folded profile with a template changes the prefactor by a factor of order unity that depends on the pulse shape~\cite{Cordes:2010fh}.} Terzan 5 is timed today with the Green Bank Telescope (GBT) and MeerKAT at 1.4~GHz, the frequency at which both telescopes do most of their pulsar timing (a band that balances the brighter emission of pulsars at low frequency against the stronger dispersion and scattering). The faint Terzan 5 pulsars have residuals of tens of microseconds, consistent with this scaling~\cite{Padmanabh2024}. The GBT, with which most of the published residuals were measured, has $A_{\rm eff}/T_{\rm sys} \simeq 280~{\rm m^2\,K^{-1}}$.\footnote{The GBT L-band receiver has a gain of $2.0~{\rm K\,Jy^{-1}}$, i.e.\ $A_{\rm eff} = 2k_{\rm B}G \simeq 5500~{\rm m^2}$, and $T_{\rm sys} \simeq 20$~K~\cite{GBTdocs}.} The mid-frequency Square Kilometre Array (SKA-Mid) will reach up to $1750~{\rm m^2\,K^{-1}}$~\cite{Braun:2019gdo}, lowering the residual of each pulsar by a factor of six. The full array envisioned for a second phase (SKA2), whose scope is still to be determined~\cite{Braun:2019gdo}, aims for a collecting area of $10^6~{\rm m^2}$~\cite{Dewdney:2009tmd}; we take its sensitivity to be $10^4~{\rm m^2\,K^{-1}}$, which lowers the present residuals by a factor of 36.

Terzan 5 also introduces cluster-specific timing structure. The first spin-period derivative $\dot\nu$ is dominated by the smooth acceleration of the cluster, while the jerk $\ddot\nu$ receives comparable contributions from the mean field and from nearby stars~\cite{Prager2017}. These contributions differ from pulsar to pulsar and act as additional pulsar-specific low-frequency structure in a Shapiro search. Redback systems, in which the pulsar is eclipsed by a low-mass companion, add wandering of the orbit to this~\cite{Rosenthal:2024zzg} and have been excluded from some dynamical analyses of the cluster~\cite{Smith2024}; they would require dedicated orbital and red-noise models.

\section{Stellar Backgrounds and the Compact-Baryon Floor}
\label{sec:baryon}
A star or stellar remnant crossing the line of sight produces the same Shapiro signal as a dark object of the same mass and trajectory, so stars of Terzan 5 and of the Milky Way are a foreground to the search. We quantify this foreground by the dark-matter fraction that would produce the same number of transits, which we call the compact-baryon floor; where this floor is comparable to the projected reach, a background-free sensitivity estimate is insufficient. By Eqs.~\eqref{eq:nexp} and \eqref{eq:area}, the number of transits of a population is proportional to its column and to its mean-square transverse speed relative to the line of sight, so equating the stellar transits per logarithmic interval of mass to those of dark objects of the same mass gives\footnote{The timing-model fit cancels in the ratio of stellar and dark-matter event rates at fixed mass.}
\begin{equation}
f_{\rm floor}(M) = \sum_i \frac{\langle v_i^2\rangle}{\langle v_{\rm DM}^2\rangle}\,\frac{1}{\Sigma_{\rm DM}}\frac{d\Sigma_i}{d\ln M},
\label{eq:baryon_floor_master}
\end{equation}
where the sum runs over the stellar populations of Terzan 5 and of the disk and bulge of the Milky Way, $d\Sigma_i/d\ln M$ is the column of population $i$ per logarithmic interval of mass, and $\langle v_i^2\rangle$ its mean-square transverse speed relative to the line of sight, with $\langle v_{\rm DM}^2\rangle$ that of the dark matter. The speeds, derived in App.~\ref{app:velocities}, are about $30~{\rm km\,s^{-1}}$ for objects bound to Terzan 5, $110$--$180~{\rm km\,s^{-1}}$ for stars of the Milky Way, $340$--$370~{\rm km\,s^{-1}}$ for its neutron stars, which are kicked at birth, and $250~{\rm km\,s^{-1}}$ for dark matter. Objects bound to Terzan 5 move with the pulsars, so only the velocity dispersion of the cluster carries them across the sightlines, suppressing their transit rate by $(30/250)^2 \simeq 1/70$ relative to dark matter of the same column.

The foreground from Terzan 5 is its present-day mass function: the living brown dwarfs and main-sequence stars together with the white dwarfs and neutron stars left by the stars that have died; we omit stellar black holes, whose masses, $\gtrsim 5\,M_\odot$, lie above the range probed. Terzan 5 is old, with a dominant population of age $\simeq 12$~Gyr~\cite{Zullo:2026tz5},\footnote{A younger, metal-rich population of age $\simeq 5$~Gyr~\cite{Zullo:2026tz5} holds up to $40\%$ of the mass~\cite{Smith2024}; we neglect it, since we estimate that its inclusion would not substantially modify the foreground below $1\,M_\odot$, where the search is most sensitive.} so we treat it as a single population of that age: every star heavier than the main-sequence turnoff, $0.9\,M_\odot$ for stars of age 12 Gyr and the metallicity of Terzan 5~\cite{Smith2024,Choi:2016mist}, has become a remnant and everything lighter is still a brown dwarf or on the main sequence. Every population then follows from the initial mass function $(dn/dM)_{\rm init}$, the number of stars formed per unit mass, for which we take the bottom-light broken power law fit to globular clusters~\cite{Baumgardt2023},
\begin{equation}
\bigg(\frac{dn}{dM}\bigg)_{\rm init} \propto
\begin{cases}
M^{-0.3}, & M < 0.4\, M_\odot, \\
M^{-1.65}, & 0.4\, M_\odot \leq M < 1\, M_\odot, \\
M^{-2.3}, & M \geq 1\, M_\odot,
\end{cases}
\label{eq:imf-t5}
\end{equation}
continuous across the breaks, which leaves a single overall normalization.\footnote{It is common to also multiply the initial mass function by an exponential roll-off, e.g., $\exp[-(0.01\, M_\odot/M)^2]$, to represent a low-mass cutoff, which has yet to be detected in the field or in young clusters~\cite{Kirkpatrick2024,Luhman:2026ngc2024}; in practice, however, we are insensitive to this roll-off, so we neglect it in our analysis.}

The fate of each star is fixed by its initial mass $M_{\rm init}$:
\begin{itemize}
\item $M_{\rm init} < 0.9\,M_\odot$: the star is either a brown dwarf or on the main sequence, where it loses a negligible fraction of its mass, so the present-day mass function of these stars is $(dn/dM)_{\rm init}$ itself.
\item $0.9\,M_\odot < M_{\rm init} < 8\,M_\odot$: the star is now a white dwarf, with a mass given by the piecewise-linear initial--final mass relation of Ref.~\cite{Cummings2018},\footnote{The white-dwarf mass is $M = 0.080\,M_{\rm init} + 0.489\,M_\odot$ for $M_{\rm init} < 2.85\,M_\odot$, $M = 0.187\,M_{\rm init} + 0.184\,M_\odot$ for $2.85\,M_\odot \leq M_{\rm init} < 3.60\,M_\odot$, and $M = 0.107\,M_{\rm init} + 0.471\,M_\odot$ for $3.60\,M_\odot \leq M_{\rm init} \leq 8\,M_\odot$, extrapolating beyond the $7.2\,M_\odot$ to which the relation is calibrated~\cite{Cummings2018}.} which compresses this range into $0.56$--$1.3\,M_\odot$.
\item $8\,M_\odot < M_{\rm init} < 20\,M_\odot$: the star is now a neutron star, with a mass near $1.4\,M_\odot$ and a spread of $0.1$--$0.2\,M_\odot$~\cite{OzelFreire2016}, which we represent by a Gaussian of width $0.15\,M_\odot$.
\end{itemize}
The remnants in an interval $dM$ of present-day mass are the stars born in the interval $dM_{\rm init}$ that the initial--final mass relation maps onto it, reduced by the fraction $\eta$ of the class that the cluster retains,
\begin{equation}
\bigg(\frac{dn}{dM}\bigg)_{\rm rem} = \eta\, \bigg(\frac{dn}{dM}\bigg)_{\rm init} \frac{dM_{\rm init}}{dM},
\label{eq:remnants}
\end{equation}
with $\eta = 1$ for white dwarfs and $0.1$ for neutron stars~\cite{Dickson:2023mfp}.\footnote{White dwarfs form without a supernova and are all retained, whereas neutron stars receive a natal kick of a few hundred ${\rm km\,s^{-1}}$, the recoil of the asymmetric explosion that forms them, which far exceeds the central escape speed of Terzan 5, $\simeq 50~{\rm km\,s^{-1}}$~\cite{Prager2017}.} Remnants carry $\simeq 35\%$ of the mass of the cluster, nearly all of it in white dwarfs.

With every population proportional to the initial mass function, the one free amplitude is fixed by the column of the cluster in front of the pulsars. We take the stellar mass density of Terzan 5 to follow a King profile~\cite{king1962structure},
\begin{equation}
\rho(r) = \rho_0 \bigg(1 + \frac{r^2}{r_c^2}\bigg)^{-3/2},
\label{eq:king}
\end{equation}
with central density $\rho_0 = 1.6\times 10^6\, M_\odot\,{\rm pc}^{-3}$ and core radius $r_c = 0.16$~pc, as measured from the accelerations of the pulsars~\cite{Prager2017}. Since the position of each pulsar along the line of sight is unknown, we average the stellar column in front of it over that position with the King profile as the prior; averaged over the measured angular positions of the pulsars, this is $9.2\times 10^4\, M_\odot\,{\rm pc}^{-2}$. While this is $\sim 500$ times larger than the dark-matter column, the difference in the speed of stellar matter dramatically reduces its influence on pulsar timing searches.

The Milky Way adds the stars of its thin disk, thick disk, and bulge along the line of sight. We replace the initial mass function of Eq.~\eqref{eq:imf-t5} by that measured for the stars and brown dwarfs within $20$~pc of the Sun~\cite{Kirkpatrick2024},
\begin{equation}
\bigg(\frac{dn}{dM}\bigg)_{\rm init} \propto
\begin{cases}
M^{-0.6}, & 0.01 \leq M/M_\odot < 0.05, \\
M^{-0.25}, & 0.05 \leq M/M_\odot < 0.22, \\
M^{-1.3}, & 0.22 \leq M/M_\odot < 0.55, \\
M^{-2.3}, & 0.55 \leq M/M_\odot,
\end{cases}
\label{eq:imf-mw}
\end{equation}
extrapolating the high-mass slope beyond the $8\,M_\odot$ to which the fit is calibrated. For the thin disk we assume a constant star-formation rate over $10$~Gyr and a main-sequence lifetime of $10~{\rm Gyr}\,(M/M_\odot)^{-2.5}$, so stars below $1\,M_\odot$ are all still on the main sequence while, of the heavier ones, a fraction $(M/M_\odot)^{-2.5}$ remains and the rest have become remnants; for the thick disk and bulge we take a turnoff of $1\,M_\odot$. White dwarfs and neutron stars follow the same initial--final mass relations as in Terzan 5 and, with nothing to eject them, are all retained.

As for the cluster, the amplitude of each component is fixed by its column. We integrate the densities of the mass model of Ref.~\cite{McMillan2017}, whose parameters are listed in App.~\ref{app:mw-model}, along the line of sight, which gives $1100$, $140$, and $95~M_\odot\,{\rm pc}^{-2}$ for the thin disk, thick disk, and bulge. The total is $70$ times smaller than the column of the cluster, but its stars move $4$--$6$ times faster relative to the line of sight. Additionally, since the Milky Way is still forming stars, its main sequence extends above the turnoff of the cluster, which pushes its contribution to heavier masses. The resulting floor, summed over the cluster and the Milky Way, is the shaded region of Fig.~\ref{fig:reach}.

\section{Results}
\label{sec:results}
\begin{figure}[t]
    \centering
    \includegraphics[width=\columnwidth]{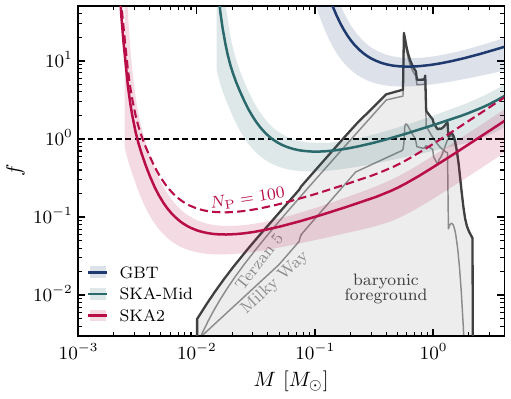}
    \caption{Projected limit on the fraction of dark matter in compact objects, $f$, from a single-event Shapiro search toward Terzan 5, for 40-yr campaigns of one 4~h visit per week, about $210$ hours of telescope time per year. Blue (GBT): current capabilities, with the published residuals treated as the uncertainty of a 4~h visit for the 16 pulsars that have one, with residuals drawn uniformly between 10 and $30~\mu$s for the other 33. Green (SKA-Mid): the same residuals divided by six, the sensitivity of SKA-Mid relative to the GBT, for 100 pulsars. Red (SKA2): the same residuals divided by 36, for 200 pulsars; the dashed red line is the Terzan 5 mix for SKA2 with the 100 pulsars of the SKA-Mid campaign. In the green and red bands, the pulsars beyond the known 49 draw their residuals from the known ones. Each band spans the timing-model fit through $\dot\nu$ (lower edge) to the fit through $\dddot\nu$ for every pulsar (upper edge); the solid line is the mix measured in Terzan 5, in which 13 of 36 pulsars require $\dddot\nu$. White timing noise is assumed. The shaded region lies below the compact-baryon foreground, whose parts from Terzan 5 and the Milky Way are shown as thin gray lines, and the dashed line marks $f = 1$.}
    \label{fig:reach}
\end{figure}

We now combine the reach of Sec.~\ref{sec:signal} with the foreground of Sec.~\ref{sec:baryon} to estimate the projected sensitivity of pulsar timing of Terzan 5 to transits of compact dark matter.

Fig.~\ref{fig:reach} compares three 40-yr campaigns, each with one 4~h visit per week ($210$~h of telescope time per year), against the compact-baryon foreground. The campaigns differ in timing precision and pulsar populations. The blue band, labeled GBT, represents current capabilities: it takes the published residuals of the 16 pulsars that have one, most of which were measured with the GBT, treated as the uncertainty of a 4~h visit, and residuals drawn uniformly between 10 and 30~$\mu$s for the other 33, combining the $\sim 20$~yr of existing timing of Terzan 5~\cite{Padmanabh2024} with 20 further years.\footnote{We treat the full 40~yr as uniformly sampled at the present precision. The archival data are sparser and noisier, so this campaign bounds from above what the existing baseline contributes.} The green band divides these residuals by six, the sensitivity of SKA-Mid relative to the GBT in the radiometer limit of Sec.~\ref{sec:radiometer}, and assumes 100 pulsars, since a deeper search of the cluster is expected to find many more. The red band assumes a telescope 36 times more sensitive than the GBT, representative of SKA2. The band and its solid line assume 200 pulsars, and the dashed line assumes 100. In the green and red bands, the pulsars beyond the known 49 are assigned residuals drawn from the known ones.

Each band is computed from Eq.~\eqref{eq:schematic-reach} with the detectable area of Eq.~\eqref{eq:area} summed over the pulsars. Its lower edge is the timing-model fit through $\dot\nu$, its upper edge the fit through $\dddot\nu$ for every pulsar, and the solid line the mix measured in Terzan 5, for which we weight the areas of the fits through $\ddot\nu$ and $\dddot\nu$ by $2/3$ and $1/3$~\cite{Prager2017}; the width of the band, a factor of about two, is the uncertainty from the timing model. We find that the present precision restricts the reach to $f > 1$, short of any physical dark-matter fraction, while SKA-Mid and SKA2 reach $f < 1$. The best limits are $f \simeq 8$ at $0.6\,M_\odot$ at the present precision, $f \simeq 0.7$ at $0.1\,M_\odot$ for SKA-Mid, and $f \simeq 0.06$ at $0.02\,M_\odot$ for SKA2. The curves assume white timing noise. Since the GWB is nearly common to the pulsars of the cluster, including it with the correlation filter of App.~\ref{app:gwb} does not meaningfully change the reach.

\section{Conclusion}
In this work, we explored the viability of pulsars in inner-Galaxy globular clusters as targets in the search for compact dark matter. Since the Shapiro delay of a transiting object accumulates along the entire photon path, the reach on the dark-matter fraction scales inversely with the column along the sightline, $f \propto 1/\Sigma_{\rm DM}$. Toward the inner Galaxy the column is an order of magnitude or more above that toward the pulsars of the timing arrays, and a cluster such as Terzan 5 enables simultaneously observing dozens of millisecond pulsars within a single telescope beam. The timing precision of pulsars in globular clusters such as Terzan 5 is in the range of $10$--$100\ \mu$s, orders of magnitude poorer than that of the pulsars used to search for gravitational-wave backgrounds. Nevertheless, the higher transit rate along the large column of the inner Galaxy can compensate.

At the published precision, even a 40-yr baseline falls roughly an order of magnitude short of $f = 1$. SKA-Mid, which lowers the timing noise by a factor of six and could significantly increase the number of known pulsars, brings the reach below $f = 1$, and SKA2 would extend it by a further order of magnitude. This effort is particularly important in light of the recent observation of a nanohertz gravitational-wave background~\cite{NANOGrav2023,EPTA2023}, which dominates the noise at the low frequencies where a long transit deposits most of its power. Since the pulsars of Terzan 5 lie within a few parsecs of one another, far less than a nanohertz wavelength, the background is nearly identical across the cluster and a correlation filter can largely remove it.

The environment that supplies the column also supplies a foreground, since a star or stellar remnant crossing a sightline produces the Shapiro delay of a dark object of the same mass. We quantified it through the compact-baryon floor, the dark-matter fraction that would produce the same number of transits, built from the present-day mass function of Terzan 5 and the stars of the Milky Way along the sightline. Although the stellar column of the cluster is $\sim 500$ times the dark-matter column, its stars share the motion of the pulsars, which suppresses their transit rate.

For compact objects, pulsar timing is at best comparable to the existing microlensing and dynamical limits~\cite{Mroz2024, Tisserand2007, Niikura2019, MACHO2001, Oguri2018, MonroyRodriguez2014, Brandt2016}. Its distinctive strength is sensitivity to {\em diffuse} substructure~\cite{Dror:2019twh,Ramani:2020hdo,Lee:2021zqw}, which is too extended to lens light but threads the same column; extending the present analysis to such structure is left to future work. Realizing these prospects calls for dedicated timing programs on the inner-Galaxy clusters with the telescopes now coming online, and for pulsar searches behind the Galactic Center, where the column, and with it the reach, is largest.

{\it Note added.}---While this work was being completed, Ref.~\cite{Cherukupalli:2026hdb} appeared, which also proposes suppressing the GWB in searches for dark-matter substructure by differencing the residuals of closely separated pulsars, such as those in dense globular clusters like Terzan 5. That work focuses on the gain from close pulsar pairs and notes that the baryonic environment of a cluster must be separated from a dark-matter signal. Here we forecast the reach of a dedicated Terzan 5 campaign to compact dark matter and estimate this baryonic foreground.

\begin{acknowledgments}
We thank Joshua Foster for useful discussions, and in particular for pointing out the importance of the higher-order timing-model parameters fitted for globular-cluster pulsars. The research of JD was supported in part by the U.S.~Department of Energy grant number DE-SC0025569.
\end{acknowledgments}

\appendix
\section{Milky Way model and relative transverse velocities}
\label{app:velocities}
\label{app:mw-model}
We model the stellar disks of the Milky Way as falling exponentially with the radius in the Galactic plane, measured from the Galactic Center, and with the height above the plane, and the bulge as a flattened power law with a Gaussian cutoff~\cite{McMillan2017}; Table~\ref{tab:mwmodel} lists the parameters. We take the Sun to lie $21$~pc above the Galactic plane~\cite{BennettBovy2019} and integrate these densities along the line of sight to Terzan 5 to obtain the columns of Sec.~\ref{sec:baryon}.
\begin{table}[tbp]
\begin{ruledtabular}
\begin{tabular}{lccc}
Component & Normalization & Scale length & Scale height \\
\hline
Thin disk & $896~M_\odot\,{\rm pc}^{-2}$ & $2.50$~kpc & $0.3$~kpc \\
Thick disk & $183~M_\odot\,{\rm pc}^{-2}$ & $3.02$~kpc & $0.9$~kpc \\
Bulge & $98.4~M_\odot\,{\rm pc}^{-3}$ & $75$~pc & -- \\
\end{tabular}
\end{ruledtabular}
\caption{Stellar components of the Milky Way model~\cite{McMillan2017}. The bulge is a power law of slope $1.8$ in a radius flattened toward the Galactic plane with an axis ratio of $0.5$, with the listed scale radius and a Gaussian cutoff at $2.1$~kpc.}
\label{tab:mwmodel}
\end{table}

The number of transits depends on the transverse velocity of the objects relative to the line of sight, which itself moves with the Earth and the pulsar. For an object at a distance $D_{\rm L}$ along the line of sight to a pulsar at a distance $D_{\rm P}$, the relative velocity is
\begin{equation}
\mathbf v = \mathbf v_{\rm L} - \frac{D_{\rm L}}{D_{\rm P}}\,\mathbf v_{\rm P} - \bigg(1-\frac{D_{\rm L}}{D_{\rm P}}\bigg)\mathbf v_{\oplus},
\label{eq:app-vrel}
\end{equation}
where $\mathbf v_{\rm L}$, $\mathbf v_{\rm P}$, and $\mathbf v_\oplus$ are the velocities of the object, the pulsar, and the Earth in the Galactic rest frame, projected onto the plane transverse to the line of sight. We average the square of Eq.~\eqref{eq:app-vrel} over the velocity distribution of each population and over the orbit of the Earth, weighting by the density of the population along the line of sight. The velocity of the Earth is the sum of the velocity of the Sun, a circular speed of $233~{\rm km\,s^{-1}}$~\cite{Evans2019} plus a peculiar velocity of $(11, 12, 7)~{\rm km\,s^{-1}}$ toward the Galactic Center, along the Galactic rotation, and toward the North Galactic Pole~\cite{Schoenrich:2009bx}, and the orbital velocity of the Earth. The velocity of a pulsar is the sum of the bulk velocity of Terzan 5, from its proper motion~\cite{VasilievBaumgardt2021} and radial velocity~\cite{Taylor2022}, and its velocity within the cluster, which objects bound to the cluster share. Table~\ref{tab:velocity_inputs} lists the mean velocity and the dispersion of each population, and Table~\ref{tab:velocity_inputs_updated} the resulting relative speeds.
\begin{table}[tbp]
\begin{ruledtabular}
\begin{tabular}{lccc}
Population & Mean velocity & Dispersion & Ref. \\
 & [${\rm km\,s^{-1}}$] & [${\rm km\,s^{-1}}$] & \\
\hline
Terzan 5 & $(-59, 68, -27)$ & $15.5$ & \cite{VasilievBaumgardt2021,Taylor2022} \\
Dark matter & $0$ & $160$; $228$, $82$ & \cite{Evans2019} \\
Thin disk stars & $239$ & $(31, 20, 11)$ & \cite{Vieira2022} \\
Thick disk stars & $225$ & $(49, 35, 22)$ & \cite{Vieira2022} \\
Thin disk brown dwarfs & $239$ & $(19, 11, 8)$ & \cite{Hsu2024} \\
Disk white dwarfs & as stars & $(31, 24, 20)$ & \cite{Kilic2019} \\
Bulge & $0$ & $117$, $102$ & \cite{Spaenhauer1992} \\
Neutron-star kick & as parent & $229$ & \cite{Igoshev2020} \\
\end{tabular}
\end{ruledtabular}
\caption{Velocities of the populations in the Galactic rest frame. The mean velocity of Terzan 5 is given in Galactic Cartesian components and that of the disks is their rotation speed; dispersions are isotropic where a single number is given and otherwise in the radial, azimuthal, and vertical directions. The dark-matter distribution is an isotropic component carrying $80\%$ of the density and a radially biased component with the listed radial and tangential dispersions carrying the rest, both truncated at an escape speed of $528~{\rm km\,s^{-1}}$. The bulge dispersions in Galactic longitude and latitude follow from proper-motion dispersions of $3.2$ and $2.8~{\rm mas\,yr^{-1}}$ at $7.7$~kpc. Neutron stars outside the cluster receive the listed isotropic kick in addition to the dispersion of their parent population.}
\label{tab:velocity_inputs}
\end{table}
\begin{table}[tbp]
\begin{ruledtabular}
\begin{tabular}{lcc}
 & \multicolumn{2}{c}{Relative speed [${\rm km\,s^{-1}}$]} \\
Population & Stars & Neutron stars \\
\hline
Terzan 5 & 30 & 30 \\
Thin disk & 110--120 & 340 \\
Thick disk & 110 & 340 \\
Bulge & 180 & 370 \\
\end{tabular}
\end{ruledtabular}
\caption{Root-mean-square transverse speed of each stellar population relative to the line of sight to Terzan 5. The first column applies to brown dwarfs, main-sequence stars, and white dwarfs. The corresponding speed for dark matter is $250~{\rm km\,s^{-1}}$.}
\label{tab:velocity_inputs_updated}
\end{table}

\section{The timing-model fit}
\label{app:timing-fit}
We evaluate the coefficient $C(t_0,\tau)$ of Eq.~\eqref{eq:snr-c} at the observation times. We write $R = -2GM\,s$, with $s_i = \ln[\tau^2 + (t_i - t_0)^2]$ at the $N = T/\Delta t$ observations, and collect the terms of the timing model in the design matrix $X_{ik} = t_i^{\,k}$, $k = 0, \dots, n$. The fit subtracts the least-squares polynomial $X\theta^*$, where $\theta^* = (X^TX)^{-1}X^Ts$ is the shift that the transit induces in the spin parameters. This leaves the post-fit residual $s_\perp = (1-\Pi)\,s$, with
\begin{equation}
\Pi \equiv X(X^TX)^{-1}X^T .
\end{equation}
Since $\Pi$ is symmetric and squares to itself, so does $1-\Pi$, and $C$ in Eq.~\eqref{eq:snr-c} becomes
\begin{equation}
C(t_0,\tau) = \frac{4}{N}\, s^T (1-\Pi)\, s .
\label{eq:c-general}
\end{equation}

\begin{figure}[t]
    \centering
    \includegraphics[width=\columnwidth]{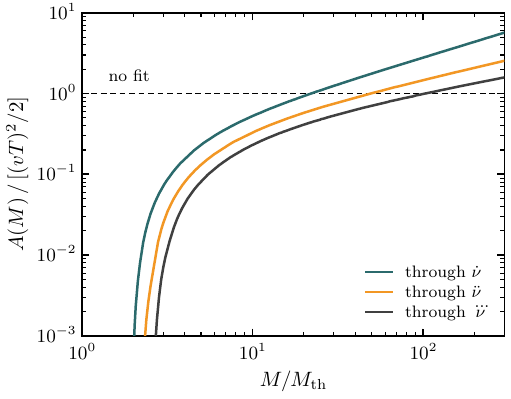}
    \caption{Detectable area of Eq.~\eqref{eq:area} for a single sightline, in units of $(vT)^2/2$, the area of the transits contained in the campaign, for the timing-model fits through $\dot\nu$, $\ddot\nu$, and $\dddot\nu$. The dashed line is the area without a fit, for which every contained transit above $M_{\rm th}$ is detectable.}
    \label{fig:area}
\end{figure}

Evaluating Eq.~\eqref{eq:c-general} on a grid of $t_0$ and $\tau$ that extends beyond the campaign, we count the trajectories that satisfy Eq.~\eqref{eq:detect} to obtain the detectable area of Eq.~\eqref{eq:area}. Fig.~\ref{fig:area} shows it in units of $(vT)^2/2$, the area of the transits contained in the campaign, for the fits through $\dot\nu$, $\ddot\nu$, and $\dddot\nu$.

\section{Gravitational-wave background}
\label{app:gwb}
\begin{figure}[t]
    \centering
    \includegraphics[width=\columnwidth]{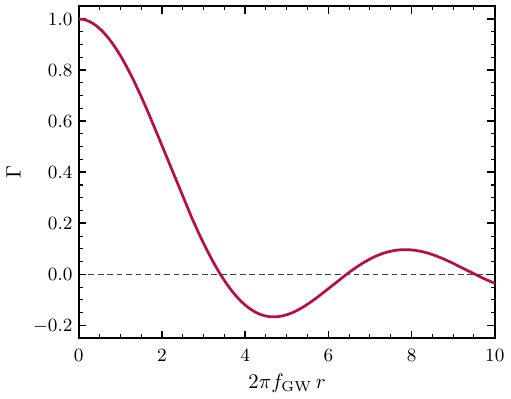}
    \caption{Correlation of the pulsar terms of the GWB between two pulsars in the same direction, Eq.~\eqref{eq:rp}, as a function of their transverse separation $r$ in units of the reduced wavelength.}
    \label{fig:coherence}
\end{figure}
After the timing-model fit, the power of a transit sits in the lowest few harmonics of the campaign, where the GWB exceeds the white timing noise. At a frequency $f_{\rm GW}$, we take the power spectrum of the timing residuals induced by the GWB to be
\begin{equation}
\mathcal{P}_{\rm GWB}(f_{\rm GW}) = \frac{A_{\rm{GWB}}^2}{12\pi^2}\, \bigg(\frac{f_{\rm GW}}{1~{\rm yr}^{-1}}\bigg)^{-13/3}\,{\rm yr}^{3},
\label{eq:pgwb}
\end{equation}
with a strain amplitude $A_{\rm{GWB}} = 2.4\times 10^{-15}$~\cite{NANOGrav2023}, and that of the white noise to be $2\Delta t\, t_{\rm RMS}^2$. For a 40-yr campaign of weekly visits with $3\ \mu$s residuals, the GWB exceeds the white noise by a factor of about $10^3$ at the lowest frequency, $f_{\rm GW} = 1/T$.

The GWB is correlated between the pulsars. The delay induced at the Earth is common to all pulsars~\cite{HellingsDowns1983, Anholm2009}, while the delays induced at two pulsars at the same distance are correlated only when their transverse separation is less than a wavelength. The covariance between the residuals of pulsars $a$ and $b$ at frequency $f_{\rm GW}$ is
\begin{equation}
\sigma_{ab} = 2\Delta t\, t_{\rm RMS}^2\,\delta_{ab} + \frac{1}{2}\mathcal{P}_{\rm GWB}(f_{\rm GW})\,\big[1 + \Gamma(f_{\rm GW}, r_{ab})\big],
\label{eq:covariance}
\end{equation}
where $r_{ab}$ is the transverse separation of the two pulsars and the correlation of the pulsar terms~\cite{Mingarelli:2014xfa}, evaluated for two pulsars in the same direction, is
\begin{equation}
\Gamma(f_{\rm GW}, r) = \frac{3}{8} \int_{-1}^{1} d\mu\; (1 - \mu)^2\, J_0\Big(2\pi f_{\rm GW}\, r \sqrt{1 - \mu^2}\Big),
\label{eq:rp}
\end{equation}
with $\mu$ the cosine of the angle between the direction of the gravitational wave and the line of sight and $J_0$ a Bessel function. Eq.~\eqref{eq:covariance} differs from the Hellings--Downs correlation because the pulsars lie in a single direction and, at the few lowest harmonics where the GWB exceeds the white noise, within a wavelength of one another: the wavelength at $f_{\rm GW} = 1/T$ is $12$~pc for $T = 40$~yr, while the median separation of the Terzan 5 pulsars is $0.5$~pc. The Hellings--Downs curve is the correlation of the Earth term between pulsars in different directions, with the pulsar terms dropped because distant pulsars are many wavelengths apart. Here the Earth term takes its zero-separation value, the $1$ in the bracket, and the pulsar terms remain correlated through $\Gamma$, shown in Fig.~\ref{fig:coherence}. For a transit with residual $R_a(t)$ in pulsar $a$, the significance of Eq.~\eqref{eq:snr-c} is replaced by
\begin{equation}
{\rm SNR}^2 = \sum_{k} \sum_{a,b} \tilde R_a^*(f_k)\, \sigma^{-1}_{ab}(f_k)\, \tilde R_b(f_k),
\label{eq:snr-filtered}
\end{equation}
where $\tilde R_a(f_k)$ is the Fourier amplitude of the residual at the frequencies $f_k = k/T$ of the campaign.\footnote{We define $\tilde R_a(f_k) \equiv \frac{2}{\sqrt{T}}\int_0^T dt\, R_a(t)\, e^{-2\pi i f_k t}$ and sum over $k \geq 1$.}
\bibliographystyle{apsrev4-2}
\bibliography{biblio}

\end{document}